\documentclass[a4paper,12pt]{article}
\usepackage{latexsym,amssymb,amsfonts,epsfig,graphicx,color,axodraw}
\newcommand{\La}{\ensuremath{\mathcal L}}
\newcommand{\be}{\begin{equation}}
\newcommand{\ee}{\end{equation}}
\title{Gravitational Coleman-Weinberg Potential and It's Finite Temperature Counterpart}
\author{Srijit Bhattacharjee$^a$, Parthasarathi Majumdar$^b$\\
$^a$1)Astroparticle Physics \& Cosmology Division,\\ Saha Institute of Nuclear Physics, Kolkata- 700064, India.\\2)Discipline of Physics, \\Indian Institute of Technology Gandhinagar\\ Ahmedabad, Gujarat- 382424, India.\\
$^b$Department of Physics,\\ Ramakrishna Mission Vivekananada University\\
Belur Math, Howrah 711202, India.}

\date{}
\begin{document}
\maketitle

\begin{abstract}
Coleman-Weinberg (CW) phenomena for the case of gravitons minimally coupled to massless scalar field is studied. The one loop effect completely vanishes if there is no self interaction term present in the matter sector. The one loop effective potential is shown to develop an instability in the form of acquiring an imaginary
part, which can be traced to the tachyonic pole in the graviton propagator. The finite temperature counterpart of this CW potential is computed to study the behaviour of the potential in the high and low temperature regimes with respect to the typical energy scale of the theory. Finite temperature contribution to the imaginary part of gravitational CW potential exhibits a damped oscillatory behaviour; all
thermal effects are damped out as the temperature vanishes, consistent with the
zero-temperature result. 

\end{abstract}

\section{Introduction}

Physics at Tev scale appears to be adequately described by the standard strong and electroweak theory. The detection of Higgs boson at LHC in July 2012 is important because Higgs mechanism is responsible for giving the masses to the elementary particles. Although we are still in the dark about the origin of the Higgs potential. We invoke a tachyonic mass term in the Higgs potential to create an instability which plays a key role on providing masses to the gauge bosons. How this particular form of the Higgs potential has arisen has not been understood satisfactorily yet. However, there is indeed a mechanism called the Coleman-Weinberg (CW) mechanism \cite{CW} where such a potential may be generated radiatively, starting with either a {\it vanishing potential}, or at least a potential without an instability.  In the Coleman-Weinberg  
mechanism, perturbatively generated infrared instabilities in the massless bosonic loops get stabilized, generating masses for the vector bosons radiatively. The first instance of the CW mechanism was demonstrated for the case of massless Klein-Gordon field with a self interacting potential and for massless scalar electrodynamics \cite{CW}. This effect  was shown to be present even for electroweak theory. However, this wonderful idea is unfortunately not useful for phenomenology as the mass for Higgs turns out to be smaller than 125 Gev. 

The functional methods to study radiative correction of classical potential was introduced by Schwinger \cite{Schwinger} and developed by Jona-Lasinio \cite{JL}. After the seminal paper by Coleman and Weinberg, and the development of the the loop expansion \cite{Jackiw}, effective potential has become almost indispensable in the discussion of vacuum instability. There are many applications of CW potential in theoretical physics. CW potential has been applied to compute lower mass bound of Higgs from the requirement of stability of electroweak vacuum. The minimum Higgs mass was calculated via Coleman-Weinberg mechanism to have the result consistent with the standard hot Big Bang model. Different methods of calculation of effective potentials especially the renormalization 
group improved version can be found in \cite{Sher}. In \cite{Sher}, finite temperature corrections to the effective potential, calculation of tunneling rates and the nature of cosmological phase transitions are also discussed. These results are then applied to the standard model to derive stringent bounds on Higgs and fermion passes. Another noteworthy application of CW theory is in the inflationary phase of early Universe. It was shown that a straightforward extension of the minimal $SU(5)$ Higgs system yields a satisfactory inflationary scenario where the inflation starts at the top of the CW potential. A scalar field that transforms as a singlet under the unifying gauge symmetry is all that needs to be added \cite{Shafi-Vil}.  Recently it has been reported that predictions for a class of realistic inflation models based on a quartic CW potential for a gauge singlet inflaton field agrees very well with the WMAP data \cite{Shafi-Seno}. 

Similarly in spirit one can ask the question whether any similar phenomenon occurs when spin-2 graviton is minimally coupled to a massless Higgs field. It may not have practical applications as in the case for standard model but it may reveal certain interesting features of unique nature of gravity as a fundamental interaction. In this paper we examine this question critically and compare it with the results known from standard models. We show, when gravitons couple to massless scalar field backgrounds which are
spacetime independent, a similar instability appears, with the effective one-loop graviton propagator acquiring a tachyonic pole. This, in turn, leads to the
appearance of an imaginary contribution in the one-loop effective action
for a wide class of theories involving graviton and scalars, when evaluated using the Euclidean path integral saturated at a saddle point characterized by a flat Euclidean metric and a constant scalar background. Thus the gravitational analogue of CW potential suffers additional instability compared to it's standard model counterpart. The possibility of a instability in flat spacetime due to quantum graviton fields has been discussed in the literature \cite{Smolin,Cho,Odintsov} earlier. However, to the best of our knowledge, this instability in the effective potential has not been explicitly traced to a tachyonic excitation in the graviton correlator.

The appearance of imaginary part in the effective potential in Higgs-graviton theory is possibly due to the backreaction of the classical background field (Higgs) on spacetime. When background field method is employed in path integral for graviton-Higgs theory the classical Einstein equation of motion is not satisfied by Minkowski space. In fact, for constant scalar background, the potential term (assuming positive) acts as a cosmological constant. This indicates that the vacuum should have to be taken as de Sitter space. However, calculating scalar effective potential in a de Sitter background would involve many difficulties. There is no well-defined propagator for a massless, minimally coupled scalar field in de Sitter symmetric states \cite{Allen1, Allen-Folacci}. This means that a perturbative calculation will break down which causes difficulties in computing rigorous quantum corrections for the de Sitter space. At present, whether this infrared divergence is physical entity or not is not clear.

In this paper we will assume that the unstable vacuum decays into some stable vacuum and unitarity is restored. Next, we may ask what happens
when the graviton-Higgs system is made to interact with a heat bath ?
The thermal contribution of the one loop effective potential is important to investigate, to ascertain whether the zero temperature
instability is reinforced or weakened. Doubtlessly, the result of this
assay will have implications for inflation and perhaps also for the
electroweak phase transition in the early Universe. The recent
discovery of a 125 Gev scalar boson at CERN  lends special credence to
theories with gravitons interacting with Higgs fields vis-a-vis their
implication for various instabilities in the early Universe
\cite{Kibble,Linde}. In this paper we also explore the interplay of thermal
vs Higgs/graviton induced instability in the finite temperature through
computation of the one loop graviton-Higgs effective potential. It is
found that the effect of constant scalar background is being amplified
in the high temperature limit of Higgs-graviton effective
potential. The low temperature limit, on the other hand,  
shows a rather interesting behaviour : in the physically relevant region the temperature dependent imaginary part oscillates with a damping amplitude. This oscillation may be a reminiscence of the instability of flat background under perturbation in presence of interaction between gravitons and thermally excited matter fields \cite{Gross}. 

The plan of the paper as follows : In the next section we briefly outline the computation of the one-loop effective potential of graviton-Higgs theory, where the instability is exhibited, and its presence traced to the tachyonic pole in the
graviton propagator. There is also a comparison of the nature of the
one-loop effect between gauge-Higgs and graviton-Higgs theory. This is
followed in section 3 by a detailed discussion on the one-loop effective potential at
finite temperature. The infrared limit describes an interesting
situation exhibiting an instability due to the temperature dependent
contribution to the effective potential developing an additional
imaginary part over and above the one in the $T=0$ limit. Various
aspects of the finite temperature effects are discussed. The concluding section contains a critical look at our results and future outlook.    

\section{Coleman-Weinberg potential for graviton-Higgs theory}

Here, we calculate the Coleman-Weinberg potential for a theory where 
gravity is coupled to a Higgs field minimally. We start with the 
the Euclidean path integral,
\be Z=\int {\cal D}g {\cal D} \chi e^{-S_E}=e^{-W}\label{PI}
\ee
where $S_E$ is the Euclidean action for the full theory and $\chi$ is any generic field. However, since the path integral has configurations which are identical under the diffeomorphic transformations we have to impose gauge condition to integrate over gauge inequivalent configurations.  

The Lagrangian (for positive definite Euclidean metric) for gravity minimally coupled to scalar field is,
\be\sqrt{g}{\cal L}=-\frac{1}{\kappa^2} R +\sqrt{g}\frac{1}{2}g^{\mu
\nu}\partial_{\mu}\phi\partial_{\nu}\phi +\sqrt{g}V(\phi)\label{Lag}\ee
The one-loop effective potential is obtained from (\ref{PI}) employing loop expansion technique \cite{Jackiw} and setting the background scalar field to be a constant. It is sufficient to expand the terms in the full Lagrangian upto quadratic in fluctuating fields to get the one-loop effective potential.

The gauge fixing Lagrangian is, 
\be {\cal
L}_{gf}=\frac{1}{2}\left[\partial_{\mu}h^{\mu\nu}-\frac{1}{2}\partial^{
\nu}h\right]^2.
\label{gf}\ee

The ghosts decouple in this gauge and don't contribute to the effective action. Retaining terms upto quadratic in fluctuating field $\Phi$ we get the Lagrangian relevant for one-loop {EP} effective potential
\begin{eqnarray}
 \La^{1}&=&\nonumber{1 \over 2}\Phi(-\Box_{E}+V^{''})\Phi+{1 \over 4}
h_{\mu
\nu}(-\Box_{E}-\kappa^2V)h^{\mu \nu}\\&-& {\kappa \over 2} h V^{'}(\phi_0)\Phi-{1
 \over 8}h\left[(-\Box_{E})-\kappa^2
V \right]h   \label{Lagrel}
\end{eqnarray}

Inserting the above Lagrangian into the path integral (\ref{PI}) and integrating over scalar fluctuations $\Phi$ we get an effective action which contains a propagator corresponding to the operator $(-\Box_{E}+V^{''})$, coming from the bilinear interaction term proportional to $h\Phi$. This effective action $\Gamma$, is now quadratic in $h_{\mu\nu}$.
\be e^{-\Gamma^{1}[\phi_0]}=e^{-V(\phi_0)-\frac{1}{2}TrLog[-\Box_{E}+V^{''}]}\int {\cal D}h_{\mu\nu}\,e^{\,\int -\frac{1}{2}h_{\mu\nu}M^{\mu\nu\alpha\beta}h_{\alpha\beta}}\label{5}\ee
To evaluate the functional integral of (\ref{5}) we write the expression $h_{\mu\nu}M^{\mu\nu\alpha\beta}h_{\alpha\beta}$ in a compact form,
\[\Psi_iM_{ij}\Psi_j\]
where $\Psi_i$ $(i=1,2,..,10)$ represents ten independent components of $h_{\mu\nu}$ \cite{'t Hooft, Veltman}. The components of the ten dimensional vector $\Psi_i$ are related to the graviton field tensor components as follows:
 \begin{eqnarray}
 \Psi_i &=& h_{ii}~,~i=1,...4 \nonumber \\
 \Psi_5 = h_{12}~ ,~ \Psi_6 &=& h_{13} ~,~\Psi_7= h_{14} \nonumber \\
 \Psi_8 = h_{23} ~,~ \Psi_9 &=& h_{24}~,~\Psi_{10} = h_{34} \label{map}
 \end{eqnarray}
The matrix elements of $M$ can now be easily obtained from (\ref{map}). The matrix takes a simple form in the chosen gauge. The lower $6 \times 10$ part of the matrix is diagonal, with each of them have same entry $k^2-\kappa^2V$. The upper $4 \times 4$ part is a symmetric matrix with diagonal entries $(-k^2 + \kappa^2 V)/4 + \frac{V^{'2}\kappa^2}{(k^2 - V^{''})}$ and off-diagonal entries are $-(-k^2 + \kappa^2 V)/4 + \frac{V^{'2} \kappa^2}{(k^2 - V)}$.

The eigenvalues for the matrix $M$ are,
\begin{eqnarray}\lambda_i&=&k^2-\kappa^2V \,\,;(1 \leq i \leq 6)\nonumber \\ 
\lambda_i&=&\frac{1}{2}(k^2-\kappa^2V)\,\,;(7 \leq i \leq 9)\nonumber \\
\lambda_{10}&=&-\frac{1}{2}\left[\frac{(k^2-\kappa^2V)(k^2+V^{''})
+2\kappa^2V^{' 2}}{k^2+V^{''}}\right]\end{eqnarray}
The eigenvalue for the operator coming from the quadratic part of the scalar field is given by,
\be \lambda_{\phi}=k^2+V^{''}\ee
The one loop effective potential now formally can be expressed as
\be V_{eff}^{1}=\frac{1}{2}TrLog[M]\,+\frac{1}{2}TrLog\lambda_{\phi}, \ee
while in terms of the momentum integrals (functional trace) the effective potential is given by,
\begin{eqnarray}
 V_{eff}^1 &=& V + {9\over2}\int \frac{d^4k}{(2\pi)^4}\ln{(k^2-\kappa^2V)} \nonumber \\
 &+& {1\over2}\int \frac{d^4k}{(2\pi)^4}
\ln{\left[k^4+\,(V^{''}-\kappa^2V)k^2+\kappa^2(2V^{'2}-VV^{''})\right]} \label{mom_rep}
\end{eqnarray}
By inspection it is clear here that in the infrared limit ($k\rightarrow0$) the momentum integrals have nonanalytic behaviour. Evaluating the momentum integrals with a cut-off $\Lambda$ the one loop effective potential is given by,

\small{\begin{eqnarray}
 V_{eff}^1(\phi_0)= &V&\nonumber+\frac{9}{32\pi^2}
\left[\frac{\kappa^4V^2}{2}\left(\ln{\frac{
\kappa^2V}{\Lambda^2}}-\frac{1}{2}\right)-
\kappa^2V\Lambda^2\right]-\frac{i 9\kappa^4V^2}{64\pi}\\
&+&\nonumber\frac{1}{32\pi^2}\left[(V^{''}-\kappa^2V)
\Lambda^2+\frac{a^2-2b}{4}\left(\ln{\frac{b}{\Lambda^4}}-1\right)\right]
\\&+&\frac{a\sqrt{a^2-2b}}{64\pi^2}\ln{\left[\frac{a+\sqrt{a^2-4b}}{a-\sqrt{
a^2-4b}}\right]} ~\label{effpot}
\end{eqnarray}} 
%
%
where

 {\begin{eqnarray*}a&=&V^{''}-\kappa^2 V \\
b&=&\kappa^2(2V^{' 2}-V V^{''})
\end{eqnarray*}}
$\Lambda$ is a momentum cut-off which we can set to be equal to Planck mass $M_{Planck}$.
The source of the imaginary part in the last term of the first line of eq. (\ref{effpot}) is the infrared sector of the graviton scalar mode-Higgs interaction : the relative sign of the graviton-Higgs couplings actually gives rise to this. 

As already mentioned, the appearance of the imaginary part can be traced to the effective graviton propagator of this theory. If we write down an effective linearized equation of motion for the graviton field from the quadratic part of the gauge fixed Lagrangian (here we go back to Lorentzian signature to evaluate the graviton propagator), we get an equation 
\be {\cal I }^{\alpha\beta,\mu\nu}h_{\mu\nu}=\kappa T^{\alpha\beta}
~, \label{eom} \ee
where $T^{\mu\nu}$ contains interaction terms containing
appropriately contracted products of terms {\it linear} in the scalar and
graviton fluctuation fields. The operator ${\cal I}^{\mu\nu,\alpha\beta}$ can be extracted from the bilinear effective
Lagrangian. We write it here in Fourier space

\be {\cal I}^{\mu\nu\alpha\beta}=(-k^2+\kappa^2 V){\cal O}^{\mu\nu,\alpha\beta}\label{operator}\ee

Now we can invert this operator to get the propagator using same gauge fixing as (\ref{gf}) ,
\be
D_{\mu\nu\alpha\beta}(k) =
{ \eta_{\mu\alpha} \eta_{\nu\beta} +
\eta_{\mu\beta} \eta_{\nu\alpha} - \eta_{\mu\nu} \, \eta_{\alpha\beta} 
\over k^2 -\kappa^2 V(\phi_0)}
\label{propagator}
\ee

Clearly, the poles of the propagator are at $k_0=\pm\sqrt{{\bf k}^2-\kappa^2 V}$ which give rise tachyon in the infrared limit for positive definite potential terms! It is clear from the expression (\ref{propagator}) that the gravitational
backreaction may not be neglected because the characteristic
instability time $|k|^{-1}$ is just of the order the Hubble time
($\kappa^2V)^{-1/2}$.  Thus,
the correct way of studying this problem would be to consider
small perturbations in the de Sitter background (for positive
V$(\phi_0)$). But then it is known that the de Sitter space-time is
only marginally stable with respect to small perturbations. 

One may think that the logarithmic terms in the second and third lines of the eq. (\ref{effpot}) may give rise imaginary contributions also. Indeed this could happen in some cases. However, this is not possible for any monomial potential with positive coefficient. The most obvious example of this kind is $\lambda \phi^4$. It is easy to see that $b$ is positive for this case and since the terms which are Planck-suppressed always dominated by unsuppressed ones both $a$ and $a^2-4b$ are positive so long as $\phi_0 < M_{Planck}$. These conditions appropriately rule out the possibility of any imaginary contribution from other terms of eq. (\ref{effpot}). However, if we don't restrict the potential to be of this particular form then for positive $V$ the sufficient condition for not getting any additional imaginary part from the logarithmic terms in the effective potential reduces to $a,b>0$.

Similar results have been obtained in \cite{Smolin,Cho} etc. In
a related work, Fradkin et al \cite{Fradkin} have shown that 
for a gauged supergravity theory one of the modes in the spectral decomposed one-loop operators in de Sitter background contains negative modes. The appearance of imaginary part in one loop effective action was also reported by Odintsov for $SU(5)$ GUT theory in de Sitter background \cite{odin}. In some higher derivative gravity
theories, with non-minimal coupling to scalar fields, similar imaginary
terms in the effective potential have also been observed \cite{Cho-hd,bhatt-chatt}.

It is worthwhile to mention here about a striking similarity between this instability and the Jean's instability, considered as one of the driving mechanism of structure formation in early Universe. In a classic work, Gross et. al. \cite{Gross} consider a gas of gravitons in thermal contact at finite spatial volume and interacting with thermally excited fermions and gravitons. In the one particle irreducible (1PI) Green's function, the graviton is shown to acquire an imaginary mass leading to
a tachyonic instability. In fact, gravitons interacting with thermally excited scalars, the longitudinal part of graviton $h_{00}$ develops a mass term due to thermal fluctuations \cite{Kikuchi}. The mass induced in the 1PI Green's function in this case is
\be m_g^2=-{4\over5}\pi^3GT^4\label{mass1}\ee
The presence of a heat bath as a source for
inducing thermal fluctuations is crucial in this work, as is evident
from the fact that the induced mass has power law dependence on
the temperature. 

Here, in this paper we have found that even at zero temperature,
when gravitons couple to massless scalar field backgrounds which are
spacetime independent, a similar instability appears, with the effective one-loop graviton propagator acquiring a tachyonic pole. The pole term in the propagator is proportional to $\phi_0^4$. Thus $\phi_0$ seems to play the role of a heat bath! This instability , in turn, leads to the
appearance of an imaginary contribution in the one-loop effective action. This is true 
for a wide class of theories involving
graviton and scalars, when evaluated using the Euclidean path integral
saturated at a saddle point characterized by a flat Euclidean metric and
a constant scalar background. This implies that graviton
fluctuations coupled to constant scalar field background
at $T=0$ in flat spacetime plays a role similar to gravitons in a finite 
temperature heat-bath inducing an instability in flat spacetime.  

An interesting feature of the one-loop effective potential is that the effect completely
disappears if the classical Higgs potential is set to zero. This is in 
contrast to the flat spacetime gauge field theories where a minimally
coupled Higgs field {\it generates} an effective 
potential perturbatively, even if the classical potential
vanishes. This is because in Higgs-graviton theory the absence of 
a classical Higgs potential, the Higgs field has no other coupling to
the graviton field when expanded around a constant vacuum value. In
standard electroweak theory in flat spacetime, in contrast, the
classical Lagrangian has Higgs-gauge field
seagull terms which lead to the one loop effective potential
\cite{sb-pm,sb-pmmg13} even in the absence of a classical potential. This does
not happen in perturbative quantum 
gravity since there is no such interaction for a constant Higgs background. This illustrates the subtle difference between general coordinate invariance and gauge invariance.

\section{Effect of Finite Temperature}

The appearance of an imaginary part in the zero temperature one-loop
effective potential prompts us to investigate the situation for the
finite temperature counterpart of the theory. We have already shown a tachyonic pole in the one loop graviton
self-energy leads to an instability in the
theory. The issues addressed in this section are : (a) how this
instability manifests in the thermal CW potential, and (b)
if there are additional {\it imaginary} temperature-dependent
contributions at one loop, whether these contributions neutralize the
zero-temperature imaginary part of the effective potential found in
the last section, or {\it enhance} it.  

From (\ref{effpot}) we can write down the expression for the one-loop
effective potential in momentum space in a slightly modified form,
\be V_{eff}= V + {9\over2}\int \frac{d^4k}{(2\pi)^4}\ln{(k^2-\kappa^2V)}\,+\,\sum_{i=1}^2{1\over2}\int \frac{d^4k}{(2\pi)^4}
\ln{\left[k^2+A_i\right]}\label{mom_rep1}\ee
with $A_i$'s are root of the quartic equation $k^4+ak^2+b=0$ where $a=V^{''}-\kappa^2V$
and $b=\kappa^2(2V^{' 2}-V V^{''})$. 

To obtain finite temperature effective potential we have to shift the momentum integrals of (\ref{mom_rep1}) by
\begin{eqnarray*}
\int d^4k\,&\rightarrow&\,T\sum_n\int d^3{\bf k}\,\nonumber \\ k&\rightarrow&(2\pi nT,{\bf k})
\end{eqnarray*}
Thus now, the finite temperature counterpart of Euclidean momentum integrals become \cite{LeBellac,Martin},
\be V_{eff}=V_0\,+\,{1\over2}T\sum_{i=1}^{2}\sum_{n=-\infty}^{\infty}\int \frac{d^3{\bf k}}{(2\pi)^3}\ln({\bf k}^2+4
\pi^2n^2T^2+\,A_i)\,+\,{9\over2}T\int \frac{d^3{\bf k}}{(2\pi)^3}\,\ln({\bf k}^2+4 \pi^2n^2T^2-\kappa^2V)\ee
We can represent the above integrals in a general form, 
\be I(t,u)=\frac{t^{1\over2}}{2\pi}\,\sum_{n=-\infty}^{n=\infty}\,\int\frac{d^3{\bf k}}{(2\pi)^3}\ln({\bf k}^2+tn^2+u)\ee

Here $t=4\pi^2T^2$. 

Since $I(t,u)$ is a divergent quantity we have to regularize this integral. Dimensional
regularization is most convenient to evaluate such integrals. We perform an integral transform to tackle the infinite
sum in the expression. The basic integral is,
\be I(t,u,d)=\frac{t^{1\over2}}{2\pi}\,\sum_{n=-\infty}^{n=\infty}\,\int\frac{d^3{\bf k}}{(2\pi)^3}\ln({\bf k}^2+tn^2+u)
=-\frac{t^{1\over2}}{2\pi}\sum_{n=-\infty}^{n=\infty}\frac{1}{(4\pi)^{d/2}}\,
\int_0^{\infty}d\tau \tau^{-d/2-1}e^{-\tau(tn^2+u)} ~,\label{I}\ee
where we have used the relation
\be \int \frac{d^d{\bf k}}{(2\pi)^d}\ln({\bf k}^2+tn^2+u)=-\frac{\partial}{\partial\alpha}\int\frac{d^d{\bf k}}{(2\pi)^d}
\frac{1}{({\bf k}^2+tn^2+u)^{\alpha}}\Bigg|_{\alpha=0}=-\frac{\Gamma(-{d\over2})}{(4\pi)^{{d\over2}}}(tn^2+u)^{d\over2}~,\ee
and also assumed that the $\tau$ integration has no singularities. 

To evaluate the integral (\ref{I})
we perform a large temperature expansion of the integrand. At high temperature limit i.e. for $\frac{u}{t}<<1$
we can write the sum over n as a binomial expansion in $\frac{u}{t}$
\begin{eqnarray}
 \sum_{n=1}^{\infty}(tn^2+u)^{d\over2} &=& \sum_{n=1}^{\infty}t^{d\over2}\left[n^{d}+({d\over2})\left(\frac{u}{t}\right)\frac{1}{n^{2-d}}
 \right. + \left.{1\over2}\left({d\over2}\right)\left({d\over2}-1\right)\left(\frac{u}{t}\right)^2\frac{1}{n^{4-d}}\,+O\left(\frac{u}{t}\right)^3\right] \nonumber\\&=&t^{d\over2}\left[\zeta(-d)+\left({d\over2}\right)\zeta(2-d)\left(\frac{u}{t}\right)
 + \zeta(4-d)\left({d\over4}\right)\left({d\over2}-1\right)+O\left(\frac{u}{t}\right)^3\right]
 \end{eqnarray}
 where we have used the definition of Riemann zeta function
 
\[\zeta(n)=\sum_{k=1}^{\infty}\frac{1}{k^n}~,~n>1 \]

and its analytic continuation to the region $n < 1$. 

One can now easily extract the pole part of the integral.
Defining $\epsilon=3-d$ the high temperature part of (\ref{I}) becomes,
\begin{eqnarray}
I(t,u,d-3)&=&-\frac{u^2}{16\pi^2}\left(\frac{1}{\epsilon}\right)-\frac{1}{6\pi^2}\zeta(-3)t^2-\frac{1}{4\pi^2}\zeta(-1)tu \nonumber+\frac{u^2}{32\pi^2}\ln{\frac{u^2}{M^2}} \nonumber \\ 
&-&\frac{1}{12\pi^2}u^{3\over2}t^{1\over2}+\frac{1}{32\pi^2}u^2\ln{u/t}\nonumber\\
 &-&\frac{u^2}{16\pi^2}\left(\gamma-3/4+{1\over2}\psi(3)-{1\over2}\ln{\frac{M^2}{\pi}}\right)+O(t^{-1})\label{ht1}
\end{eqnarray}
with \[\psi(x)={d\over dx}\Gamma[x]\] Here we have also introduced $M$ as an arbitrary scale of renormalization. From the above expression we can easily see that there will be imaginary contributions from some of the terms involved. If we closely inspect the possible $u$'s from eq. (\ref{mom_rep1}) this becomes clear. Apart from the
irrelevant constants and after getting rid of the pole term by a
suitable counter term we can write the effective potential at high
temperatures as,
\begin{eqnarray} V_{eff} &=& V + {1\over64\pi^2}\sum_{i=1}^2|A_i|^2\ln{\left(\frac{|A_i|}{M^2}\right)} +{9\kappa^4V^2\over64\pi^2}\ln{\left(\frac{\kappa^2V}{M^2}\right)} \nonumber \\ &+& V_{eff,im} +  V_{eff,T} \label{hT}
\end{eqnarray}
where, $V_{eff,im}$ consists of zero-temperature imaginary terms and $V_{eff,T}$ is the temperature-dependent part of $V_{eff}$.   

\begin{figure}[h]
 \begin{center}
\includegraphics[width=10cm,height=5cm]{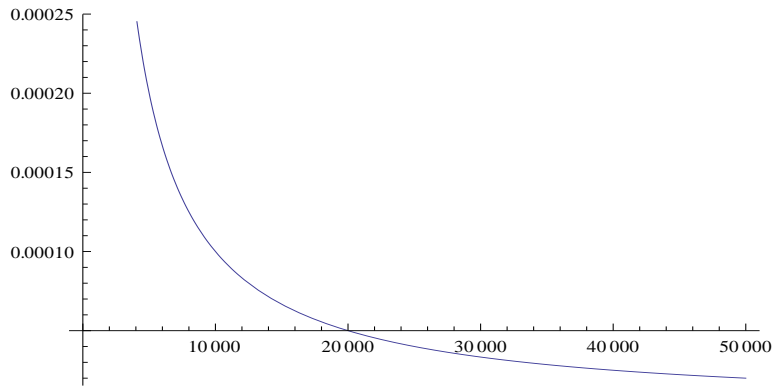}
 \end{center}
\caption{Plot of temperature dependent imaginary part of EP versus $x$, for $x<<1$}
\label{ht} 
\end{figure}

It is easy to see that the imaginary part of effective potential in this limit is
\be V^{Im}= \frac{\kappa^4V^2}{16\pi}\,+\,\frac{\kappa^3V^{3/2}}{6\pi}T\ee
This indicates, the temperature-dependent contribution to the imaginary part in fact {\it reinforces} the zero-temperature piece, thereby exacerbating the instability discussed in the last section. The plot (Fig.[\ref{ht}]) of temperature dependent imaginary contribution is simple and shows that it grows with the temperature.
where $x=\frac{\kappa V^{{1\over2}}}{T}$
It is clear that since the
dimensional pole term is proportional to $V^2$ instead of $V$ the theory is
non-renormalizable. However, our focus in not on the ultraviolet behaviour of the theory, but rather its infrared instabilities at zero and finite temperature. Thus, even if the ultraviolet divergences are tamed as conventional with appropriate counter-terms, it is obvious that the infrared instabilities will persist.

In order to ensure that the finite temperature treatment has the correct limit at vanishing temperature, one needs to consider the {\it low} temperature limit of the temperature-dependent part of $V_{eff}$. To obtain the low temperature limit of eq. (\ref{I}) we now have to use the following identity
\be \sum_{n=-\infty}^{\infty}e^{-\tau t n^2}=\left(\frac{\pi}{t\tau}\right)^{1/2}\sum_{n=-\infty}^{\infty}e^{-\pi^2n^2/t\tau}\ee
With the help of this we can write eq. (\ref{I}) as
\be I(t,u,d)=-\frac{1}{2\pi^{1/2}}\,\sum_{n=-\infty}^{n=\infty}\frac{1}{(4\pi)^{d/2}}\,
\int_0^{\infty}d\tau \tau^{-(d+3)/2}e^{-\tau t}e^{-\pi^2n^2/t\tau}\label{I1}\ee
This decomposes into two parts, one being temperature dependent and other,
zero temperature. Once again the pole
term is independent of temperature. After separating out the $n=0$ piece from the above expression we get,
\be -\frac{1}{2\pi^{1\over2}}\frac{1}{(4\pi)^{d/2}}u^{(d+1)/2}\Gamma\left(-{d\over2}-{1\over2}\right)-\frac{1}{\pi^{1/2}}\,\sum_{n=1}^{n=\infty}\frac{1}{(4\pi)^{d/2}}\,
\int_0^{\infty}d\tau \tau^{-(d+3)/2}e^{-\tau t}e^{-\pi^2n^2/t\tau} ~.
\ee

From the first term we easily extract the pole term
\be -\frac{u^2}{16\pi^2}\left(\frac{1}{\epsilon}\right)-\frac{u^2}{32\pi^2}\left(\psi(3)+\ln{4\pi\over u}\right) +O(\epsilon)\ee

We can perform the $\tau$ integration to get the temperature dependent part. The result is given in terms of a modified Bessel function \cite{Grad},
\be\int_0^{\infty}d\tau \tau^{-(d+3)/2}e^{-\tau t}e^{-\pi^2n^2/t\tau}=\left(\frac{tu}{\pi^2n^2}\right)K_2(2\sqrt{\pi^2n^2u/t})\ee
The low-temperature behaviour of the integral eq. (\ref{I1})
\begin{eqnarray}
  I(t,u,3-\epsilon)&=&-\frac{u^2}{16\pi^2}\left(\frac{1}{\epsilon}\right)-\frac{u^2}{32\pi^2}\left(\psi(3)+\ln{4\pi\over
      M^2}-\ln{{u\over M^2}}\right)\nonumber\\
  &-&\frac{u^2}{2^{1\over2}}\sum_{n=1}^{\infty}\left({t\over
      4\pi^2n^2u}\right)^{5\over4}e^{-(4\pi^2n^2u/t)^{1/2}} ~, \label{lt}
\end{eqnarray}
where for large value of the argument we have taken an asymptotic expansion for the modified Bessel function.

To analyze the low temperature behaviour of the potential we again rule out any imaginary part coming from the $A_i$'s according to the argument given in Section 2 and will only concentrate on $u=-\kappa^2V$ part. Then at Low temperature imaginary contribution for effective potential has the following form
\be \frac{\kappa^4V^2}{32\pi}+\frac{\kappa^4V^2}{2}\left(\frac{T^2}{\kappa^2V}\right)^{{5\over4}}
\sum_{n=1}^{\infty}\frac{1}{n^{5\over2}}(\cos nx-\sin nx)\label{lowT}\ee
The second term above is temperature dependent. We can approximate the sum as a integral over $n$ as $n$ goes upto infinity or we can compute the sum exactly. Performing both using MATHEMATICA we have found the behaviour of the second term of eq. (\ref{lowT}) is oscillatory with damping amplitude (Fig.[\ref{lt}]) for values of $x>0.6$ approximately and for large value of $x$ i.e. for $T\rightarrow 0$ the oscillations die away and temperature dependent imaginary part vanishes. The qualitative behaviour of the real part of this version of effective potential is almost similar to that off gauge-Higgs theories with a logarithmic term appearing at the loop level. This immediately indicates  existence of asymmetric vacua's. In principle one can minimize this potential setting $T=0$ and replace the cutoff $M$ by $\phi_{min}$ to get a gravitational counterpart of CW potential. 

\begin{figure}[h]
 \begin{center}
 \includegraphics[width=10cm,height=5cm]{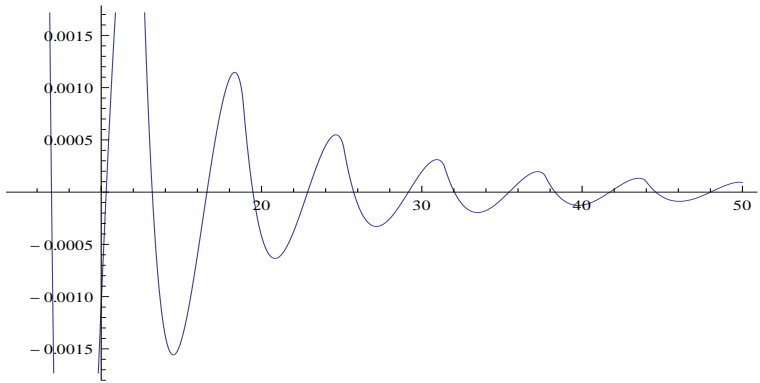}
 \end{center}
\caption{Plot of temperature dependent imaginary part of EP versus $x$, $x>>1$}
\label{lt}
\end{figure}

Just for a qualitative analysis of the high temperature counterpart of the gravitational CW potential we now ignore the imaginary parts of (\ref{hT}). If we do that then eq. (\ref{hT}) looks very similar to the usual CW potential of standard model. Now one may ask whether restoration of symmetry happens here as we higher the temperature. When $T^2\gg\phi_0^2$, the terms involving $\phi_0^2T^2$ will be dominating over $\phi_0\ln{\phi_0}$ terms in eq. (\ref{hT}). Thus there will be a complete symmetric phase after a critical temperature $T_c$ which can be estimated roughly close to $M_{Planck}10^{19}$ GeV. Thus as we increase the temperature, gravitational CW potential becomes symmetric.

\section{Conclusion}
We have shown that gravitational analog of CW potential exhibits instability. This instability is manifested as a tachyonic
mode in the one-loop propagator. A constant
scalar field background resembles a thermal bath which backreacts to the gravitons to produce the instability in the system. This infrared instability in the effective propagator may be
regarded as a graviton induced Jeans-like instability.  This instability is also manifested in the one-loop effective potential
as an imaginary term, independent of the ultraviolet cut off. The imaginary term arises from the infrared limit of the loop integrals.

We have also computed the effect of finite temperature for the
graviton-Higgs theory and compared it with the zero temperature
result. The high temperature sector involves temperature dependent
terms which adds to the imaginary contribution obtained in the zero
temperature case. The infrared sector exhibits an instability because of
imaginary contribution from both zero temperature and temperature
dependent part. Moreover it exhibits an oscillatory behaviour which
eventually gets damped out as we lower the temperature, thereby ensuring that the temperature-dependent calculation smoothly interpolates to the previous zero-temperature one-loop effective potential.   

One possible explanation of the zero-temperature instability has been given by Smolin \cite{Smolin} where it has been claimed that it might disappear if one begins with a de Sitter background spacetime instead of a Minkowski spacetime. Computing gravitational CW potential in de Sitter background may shed light on it but we know it will be difficult to compute as in a recent work by Polyakov \cite{Pol}, it has been pointed our that even de Sitter space possesses various quantum instabilities. Scalar effective potential with gravitons running into the loops around a de Sitter background will be interesting thing to study as it will have important implications in the inflationary scenario.

\noindent {\bf Acknowledgment}: We would like to thank A. Majhi, P. Byakti and C. Chakraborty for helping us to prepare this manuscript. SB expresses his thanks to Sudipta Sarkar for helpful discussions. SB would like to acknowledge the
hospitality and support of IIT Gandhinagar, Ahmedabad, where part of this work was carried out.

\end{document}